\documentclass[]{elsarticle}
\makeatletter
\def\ps@pprintTitle{%
 \let\@oddhead\@empty
 \let\@evenhead\@empty
 \def\@oddfoot{}%
 \let\@evenfoot\@oddfoot}
\makeatother
\usepackage{fancyhdr}

\usepackage{epsfig}
\usepackage{amsfonts}
\usepackage{amssymb, graphics, amsmath}
\usepackage{dsfont}
\usepackage{hyperref}
\usepackage{color}
\usepackage{pdflscape}
\usepackage{pifont}
\usepackage{graphicx} 
\usepackage{bm}
\usepackage{slashed}

\def\slashchar#1{\setbox0=\hbox{$#1$}           
  \dimen0=\wd0                                    
  \setbox1=\hbox{/} \dimen1=\wd1                  
  \ifdim\dimen0>\dimen1                           
    \rlap{\hbox to \dimen0{\hfil/\hfil}}            
    #1                                             
  \else                                          
    \rlap{\hbox to \dimen1{\hfil$#1$\hfil}}        
    /                                           
 \fi}

\newcommand{\ali}{&\!\!\!}

\newcommand{\dslash}[1]{#1 \llap{/\kern-0.5pt}}
\newcommand{\Dslash}[1]{#1 \llap{/\kern+1.5pt}}
\newcommand{\DDslash}[1]{#1 \llap{/\kern+2.3pt}}
\newcommand{\dslashh}[1]{#1 \llap{/\kern+1pt}}

\newcommand{\order}[1]{\mathcal{O}\!\left(#1\right)\!\,}

\newcommand{\bea}{\begin{eqnarray}}
\newcommand{\eea}{\end{eqnarray}}
\newcommand{\bma}{\begin{pmatrix}}
\newcommand{\ema}{\end{pmatrix}}

\makeatletter
\renewcommand*\env@matrix[1][*\c@MaxMatrixCols c]{%
  \hskip -\arraycolsep
  \let\@ifnextchar\new@ifnextchar
  \array{#1}}
\makeatother

\begin{document}

\title{Connected and Disconnected Contractions in Pion-Pion Scattering}

\author[bonn,cas]{Neramballi Ripunjay Acharya}
\author[cas,ucas]{Feng-Kun Guo}
\author[bonn,fzj]{Ulf-G. Mei\ss{}ner}
\author[sjtu]{Chien-Yeah Seng}

\address[bonn]{Helmholtz-Institut f\"ur Strahlen- und
             Kernphysik and Bethe Center for Theoretical Physics,\\
             Universit\"at Bonn, D-53115 Bonn, Germany}
\address[cas]{CAS Key Laboratory of Theoretical Physics,
            Institute of Theoretical Physics,\\ Chinese Academy of Sciences,
            Beijing 100190, China}
\address[ucas]{School of Physical Sciences,
            University of Chinese Academy of Sciences,
            Beijing 100049, China}
\address[fzj]{Institute for Advanced Simulation, Institut f\"ur Kernphysik and
           J\"ulich Center for Hadron Physics,\\ Forschungszentrum J\"ulich,
           D-52425 J\"ulich, Germany}
\address[sjtu]{INPAC, Department of Physics and Astronomy, 
           Shanghai Jiao Tong University, Shanghai 200240, Shanghai, China}

\begin{abstract}

We show that the interplay of chiral effective field theory and
lattice QCD can be used in the evaluation of so-called 
disconnected diagrams, which appear
in the study of the isoscalar and isovector channels of pion-pion scattering
and have long been a major challenge for the lattice community. By means of
partially-quenched chiral perturbation theory, we distinguish and analyze the
effects from different types of contraction diagrams to the pion-pion scattering
amplitude, including its scattering lengths and the energy-dependence of its
imaginary part. 
Our results may be used to test the current degree of accuracy of
lattice calculation in the handling of disconnected diagrams, as well as to set
criteria for the future improvement of relevant lattice computational techniques
that may play a critical role in the study of other interesting QCD matrix
elements. 

\end{abstract}

\maketitle
\thispagestyle{fancy}

%
%

%
\section{Introduction}
Quantum Chromodynamics (QCD) is well-accepted as the  fundamental field 
theory of the strong interaction. However, a direct comparison between theoretical predictions and experiments is extremely difficult at low energies (e.g. $E\ll 
1$~GeV) due to the non-perturbative nature of the theory in that regime, 
making first-principle calculations very challenging. Lattice QCD is currently the 
most promising approach to study the low-energy behavior of QCD from first 
principles. On the other hand, low-energy effective field theories (EFTs) of 
QCD are very useful in understanding the analytic behavior of the theory, but 
can also be used to make sharp predictions once the correspoding low-energy
constants have been determined. Chiral Perturbation Theory (ChPT) is an example of such kind,  based 
on the fact that chiral symmetry is spontaneously broken at the 
hadronic scale and the lowest-lying pseudoscalar mesons play the role of 
Goldstone bosons. 

Meson-meson scattering is a perfect testing ground to check the validity of different 
effective approaches to low-energy QCD. Large amounts of data on phase shifts 
have been accumulated in various channels of isospins and partial waves, 
covering a wide range of energy from $2M_\pi$
up to 1.5~GeV or higher. The quest to reproduce these phase shifts and, in 
particular, account for the contributions from resonance states, sets a great 
challenge for both the lattice and EFT communities. The low-energy
$\pi\pi$ scattering has been calculated up to two-loop order in 
ChPT~\cite{Bijnens:1995yn}, and the phenomenologically most precise and 
model-independent description comes from the dispersion-relation treatments 
using the Roy  equations matched to ChPT~\cite{Ananthanarayan:2000ht,GarciaMartin:2011cn}. On the 
lattice side,
tremendous effort has been spent over decades in the calculation of $\pi\pi$ 
scattering lengths and phase shifts on the lattice. These works cover the 
scattering at
$I=2$~\cite{Kuramashi:1993ka,Fukugita:1994ve,Li:2007ey,Sharpe:1992pp,
Gupta:1993rn,Aoki:2002in,Du:2004ib,Chen:2005ab,Beane:2007xs,Feng:2009ij,Dudek:2010ew,Beane:2011sc,Dudek:2012gj,Helmes:2015gla}, 
$I=1$~\cite{Aoki:2007rd,Feng:2010es,Lang:2011mn,Aoki:2011yj,Pelissier:2012pi, 
Dudek:2012xn,Wilson:2015dqa,Feng:2014gba,Bali:2015gji,Bulava:2016mks,Guo:2016zos,Alexandrou:2017mpi} and 
$I=0$~\cite{Kuramashi:1993ka,Fukugita:1994ve,Fu:2013ffa,Briceno:2016mjc,
Liu:2016cba,Bai:2015nea} 
channels. 

The difficulty in the lattice study of $\pi\pi$ scattering increases with the 
decrease in isospin, partly due to the differences in the involved types of quark 
contraction diagrams.  
In general, one is calculating a four-point correlation function
$\left\langle 
\mathcal{O}_{\pi\pi}(t)\mathcal{O}_{\pi\pi}^{\prime\dagger}(t')\right\rangle$ 
where each of $\mathcal{O}$ and $\mathcal{O}'$ is an interpolating field 
operator of the form $\bar{q}\Gamma^i q(\bm{x},t)\,\bar{q}\Gamma^j 
q(\bm{y},t)$, representing a state of two pions at different spacial sites. 
These correlation functions can be decomposed into distinct sets of diagrams 
according to different Wick contractions of the quark fields. In particular, 
contraction diagrams involving quark propagators starting from and ending at points 
with the same time coordinates appear to be extremely noisy, which makes the 
extraction of lattice signals very difficult. Such contractions are called 
{\em disconnected  diagrams}.
Therefore, the $I=2$ scattering amplitude that does not involve such 
contractions is relatively easy to calculate. The $I=1$ channel is more 
difficult while the $I=0$ channel was a real challenge for a long time. 
With the development of new lattice 
techniques (e.g. the implementation of all-to-all 
propagators~\cite{Neff:2001zr,Foley:2005ac,Peardon:2009gh}), the study of 
isoscalar channel becomes 
possible, but at this stage it is safe to say 
that the level of understanding is
still relatively immature compared to the $I=2$ channel. Furthermore, 
the problem of handling disconnected diagrams also appears in the lattice 
computation of many other interesting hadronic observables, such as the parity-odd 
pion-nucleon coupling~\cite{Wasem:2011zz}. It would thus be helpful to gain 
insights from the EFTs about the analytical behavior of different contraction 
diagrams so that we have a measure of how important various kinds of 
contraction diagrams could be. This would provide a gauge of how accurate 
lattice calculation  can handle such behavior. For  $\pi\pi$ scattering, 
the first attack on such a problem was performed in Ref.~\cite{Guo:2013nja}, 
where different types of 
contractions were ordered according to both the $1/N_c$  and the chiral 
expansion. However, therein the involved contractions were only computed at the
leading order (LO) of ChPT whose use is limited as  contractions involving 
more than one closed quark loops vanish at that order. 

Partially-quenched chiral perturbation theory 
(PQChPT)~\cite{Bernard:1993sv,Sharpe:1999kj,Sharpe:2000bc,Sharpe:2001fh,
Bernard:2013kwa} (for reviews, see~\cite{Sharpe:2006pu,Golterman:2009kw}) turns out 
to be an excellent candidate to fulfill the task mentioned above and 
improve the calculations in Ref.~\cite{Guo:2013nja} to higher orders in 
order to get a quantitative control of all relevant contractions. Its underlying theory,
namely the partially-quenched QCD (PQQCD), was first invented as an aid for 
lattice QCD to deal with light fermion loops. Early numerical simulations of 
lattice QCD were bothered by the difficulty to include loops of light quarks 
(which is equivalent to the computation of the fermion determinant). 
The partially-quenched approximation, namely to take the mass of quarks that appear
in closed loops (also known as ``dynamical quarks" or ``sea quarks") to be 
unphysically large, was carried out to reduce the computational effort. The 
outcome, however, has to be extrapolated to the region where the sea quark masses 
take their physical values. This is made possible by PQQCD which first separates 
the usual fermionic quarks into ``valence" and ``sea" quarks, and introduces, for 
each valence quark, a degenerate bosonic quark (known as ``ghost quark") that 
cancels all the closed-loop contribution of the valence quark\footnote{There 
also exists an alternative formulation of PQChPT based on the replica method that 
does not involve ghost quarks~\cite{Damgaard:2000di,Damgaard:2000gh}.}. With such an
arrangement, the masses of the quarks appearing in the external legs and in 
closed loops can be made different. One can go one step further to introduce 
PQChPT which is the effective field theory of PQQCD at low energy. Since the outcomes 
of PQChPT are analytic functions of sea quarks masses, one could match them to lattice 
results at unphysical sea quark masses and simply extrapolate them to the physical region. 

Computational techniques on the lattice have improved significantly nowadays and 
many calculations can be done without using the 
partially-quenched approximation. Despite that, PQChPT still continues to be 
very useful in the understanding of, and to aid, lattice calculations. The basic 
idea is that, by suitably choosing the type of quarks that appear in external 
legs, one is able to access any specific type of Wick contraction one wants, 
provided that the valence quarks are degenerate with the sea quarks. Also, when 
one considers loop diagrams, the inclusion of ghost quarks is still required so 
that the number of degrees of freedom (DOFs) that appear in the loop is the same 
as that in the original theory. This strategy has been used to study the 
connected and disconnected diagrams in, e.g., hadronic vacuum 
polarization~\cite{DellaMorte:2010aq} or the pion scalar form 
factor~\cite{Juttner:2011ur}.
In this work, we demonstrate exactly how PQChPT can be applied to distinguish 
contributions from different Wick contractions to the $\pi\pi$ scattering 
amplitude. This task has been carried out at the LO, i.e. $\order{p^2}$, 
in Ref.~\cite{Guo:2013nja} and here we want to study the complete 
next-to-leading order (NLO), i.e. $\order{p^4}$, contributions. For the sake 
of simplicity, we work in the SU(2) version of ChPT, but a generalization to 
SU(3) is straightforward.  

This work is arranged as follows: In Section 2 we review the basic setup of 
SU$(4|2)$ PQQCD and PQChPT. Section 3 contains a classification of all types of quark 
contraction diagrams that contribute to the $\pi\pi$ scattering in all isospin 
channels. In Section 4 we provide the analytic expressions of the scattering 
amplitudes contributed by each type of contraction up to NLO in PQChPT while  
in Section 5 we show some numerical results, with a special emphasis to the 
contribution from each contraction diagram to the partial wave amplitudes.  
The final conclusions are given in Section 6.  The appendices contain analytic 
expressions of scattering lengths decomposed into contributions from different 
types of contractions, as well as PQChPT in a formulation slightly different 
from the one discussed in the main text.

%

\section{Basic setup}

We will show in the next section that the simplest PQQCD that is relevant to 
our work has a flavor-SU$(4|2)$ structure. Therefore, we  start with a very 
brief review of the theory. The quarks are grouped into a fundamental representation of 
SU$(4|2)$: 
\begin{equation}
Q=\left(u\:d\:j\:k\:|\:\tilde{j}\:\tilde{k}\right)^{T}.
\end{equation}
Here, $u,d$ are the dynamical quarks, $j,k$ and $\tilde{j},\tilde{k}$ are  the
valence and ghost quarks, respectively. The PQQCD Lagrangian is given by
\begin{equation}
\mathcal{L}=\bar{Q}(i\slashed{D}-M)Q-\frac{1}{4}G^{a\mu\nu}G^a_{\mu\nu}.
\end{equation}
For our purpose in this paper, all quarks are taken to be degenerate. 
Therefore, the quark mass matrix is simply 
$M=\mathrm{diag}\left(m_q,m_q,m_q,m_q,m_q,m_q\right)$.

The massless PQQCD Lagrangian is invariant under $Q_{L/R}\to U_{L/R}Q_{L/R}$,
where $U_{L/R}\in$SU$(4|2)_{L/R}$ are elements of a special unitary
$(4|2)$ graded symmetry group. In general, an $(a|b)$-graded matrix
$A$ has the following form:
\begin{equation}
A=\left(\begin{array}{cc}
A_{1} & A_{2}\\
A_{3} & A_{4}
\end{array}\right),
\end{equation}
where $A_{1}$ ($A_{4}$) is a $a\times a$ ($b\times b$) matrix
of c-numbers while $A_{2}$ ($A_{3}$) is a $a\times b$ ($b\times a$). 
For an $(a|b)$-graded matrix, we define a supertrace as follows:
\begin{equation}
\mathrm{Str}[A]\equiv\sum_{i=1}^{a}A_{ii}-\sum_{i=a+1}^{a+b}A_{ii} \, .
\end{equation}
It is easy to show that $\mathrm{Str}$ is cyclic, namely, if $A,B\in(a|b)$,
then $\mathrm{Str}[AB]=\mathrm{Str}[BA]$.

The effective field theory of SU$(4|2)$ PQQCD at low energy is the SU$(4|2)$ 
PQChPT. Spontaneous breaking of the SU$(4|2)_L\times$ SU$(4|2)_R$ symmetry 
down to SU$(4|2)_V$ gives rise to $6^2-1=35$ Goldstone particles. They are 
represented nonlinearly by the matrix $U$ defined as:
\begin{equation}
U=\exp\left\{\frac{2i}{F_0}\sum_{a=1}^{35}\,{\phi^aT^a}\right\}~,
\end{equation}
where $\phi^a$ are the Goldstone particles and $T^a$ are the Hermitian and 
supertraceless SU$(4|2)$ generators that fulfill the following 
normalization condition:
\begin{equation}
\mathrm{Str}[T^aT^b]=\frac{1}{2}g^{ab}
\end{equation}
where $g$  is a $35$-dimensional block-diagonal matrix:
\begin{equation}
g=\mathrm{diag}\left(I_{15},-\sigma^2,-\sigma^2,-\sigma^2,-\sigma^2,-1,-\sigma^2
,-\sigma^2,-\sigma^2,-\sigma^2,-I_3\right)~,
\end{equation}
with $I_n$  the $n$-dimensional identity matrix and $\sigma^2$ the second 
Pauli matrix.

Similar to ordinary ChPT, the PQChPT Lagrangian can be ordered according to the 
usual chiral power counting rules. At $\order{p^2}$, the Lagrangian is given by
\begin{equation}
\mathcal{L}^{(2)}= 
\frac{F_{0}^{2}}{4}\mathrm{Str}\left[(\partial_{\mu}U^{\dagger})
(\partial^{\mu}U)\right]
+\frac{F_{0}^{2}B_{0}}{2}\mathrm{Str}\left[MU^{\dagger}+UM^{\dagger}\right],
\label{eq:L2}
\end{equation}
which leads to the following form of the Goldstone boson propagator:
\begin{equation}
G^{ab}(k^2)=\frac{ig^{ab}}{k^2-(M_\pi^2)_0+i\varepsilon}~,
\end{equation}
where $(M_\pi^2)_0\equiv 2B_0 m_q$ is the pion mass squared at leading order. 
Notice that this propagator does not suffer from a double-pole sickness that 
occurs in the more general PQChPT because we take the valence and dynamical quarks to be 
degenerate (see, e.g. Ref.~\cite{Sharpe:2006pu} for more general cases where valence and dynamical quarks could have different masses).

At $\order{p^4}$, the terms in the chiral Lagrangian that are relevant to this work 
are \cite{Giusti:2008vb,Gasser:1984gg,Sharpe:2006pu}: 
\begin{eqnarray}
\mathcal{L}^{(4)} \ali = \ali
L_{0}^\mathrm{PQ}\mathrm{Str}\left[(\partial_{\mu}U^{\dagger})(\partial_{\nu}
U)(\partial^{\mu}U^{\dagger})(\partial^{\nu}U)\right] +(L_{1}^\mathrm{PQ}-
\frac{1}{2}L_ {0}^\mathrm{PQ})\mathrm{Str}
\left[(\partial_{\mu}U^{\dagger})(\partial^{\mu}U)\right]
\mathrm{Str}\left[(\partial_{\nu}U^{\dagger})(\partial^{\nu}U)\right]\nonumber
\\
\ali\ali
+(L_{2}^\mathrm{PQ}-L_{0}^\mathrm{PQ})\mathrm{Str}\left[(\partial_{\mu}
U^{\dagger} )(\partial_{\nu}U)\right]
\mathrm{Str}\left[(\partial^{\mu}U^{\dagger})(\partial^{\nu}U)\right]
\nonumber\\
\ali\ali
+(L_{3}^\mathrm{PQ}+2L_{0}^\mathrm{PQ})
\mathrm{Str}\left[(\partial_{\mu}U^{\dagger}
)(\partial^{\mu}U)(\partial_{\nu}U^{\dagger})(\partial^{\nu}U)\right]\nonumber
\\
\ali\ali 
+2B_{0}L_{4}^\mathrm{PQ}
\mathrm{Str}\left[(\partial_{\mu}U^{\dagger})(\partial^{\mu}U)\right]
\mathrm{Str}\left[U^{\dagger}M+M^{\dagger}U\right]
+2B_{0}L_{5}^\mathrm{PQ}\mathrm{Str}\left[
(\partial_{\mu}U^{\dagger})(\partial^{\mu}U)(U^{\dagger}M+M^{\dagger}U)\right]
\nonumber \\
\ali\ali
+4B_{0}^{2}L_{6}^\mathrm{PQ} \left(\mathrm{Str}\left[U^{\dagger}M
+M^{\dagger}U\right]\right)^{2} +4B_{0}
^{2}L_{7}^\mathrm{PQ}\left(\mathrm{Str}\left[U^{\dagger}M-M^{\dagger}U\right]
\right)^{2}
\nonumber
\\
\ali\ali
+4B_{0}^{2}L_{8}^\mathrm{PQ}\mathrm{Str}\left[MU^{\dagger}MU^{\dagger}
+M^{\dagger}UM^ {\dagger}U\right] \, .
\end{eqnarray}
Here, every low-energy constant (LEC) comes with a superscript 
$\mathrm{PQ}$ to stress that they are 
LECs of SU$(4|2)$ PQChPT instead of ordinary ChPT. However, a direct connection 
can be made to the SU(2) ChPT~\cite{Gasser:1983yg} by adopting the 
following set of independent LECs: 
$\{l_1,l_2,l_3,l_4,l_7,L_0^\mathrm{PQ},L_3^\mathrm{PQ},L_5^\mathrm{PQ},
L_8^\mathrm{PQ}\}$ where 
\begin{eqnarray}
l_1\ali\equiv\ali 2(2L_1^\mathrm{PQ}+L_3^\mathrm{PQ})\nonumber\\
l_2\ali \equiv\ali 4L_2^\mathrm{PQ}\nonumber\\
l_3\ali 
\equiv\ali 
-4(2L_4^\mathrm{PQ}+L_5^\mathrm{PQ}-4L_6^\mathrm{PQ}-2L_8^\mathrm{PQ} 
)\nonumber\\
l_4\ali \equiv\ali 4(2L_4^\mathrm{PQ}+L_5^\mathrm{PQ})\nonumber\\
l_7\ali \equiv\ali -8(2L_7^\mathrm{PQ}+L_8^\mathrm{PQ})\label{eq:physicalLEC}
\end{eqnarray}
When $\mathcal{L}^{(4)}$ is re-expressed in terms of this set of LECs, its 
contribution to amplitudes involving only pions as external fields will not 
depend on $L_0^\mathrm{PQ},L_3^\mathrm{PQ},L_5^\mathrm{PQ}$ and 
$L_8^\mathrm{PQ}$, because terms with these coefficients vanish due to SU(2) 
trace identities. Therefore, $\{l_1,l_2,l_3,l_4,l_7\}$ correspond to the 
physical LECs, while
$\{L_0^\mathrm{PQ},L_3^\mathrm{PQ},L_5^\mathrm{PQ},L_8^\mathrm{PQ}\}$ are the 
unphysical LECs. 
%

\section{Classification of the contraction diagrams}
Our focus is on the $\pi(p_1)\pi(p_2)\rightarrow \pi(k_1)\pi(k_2)$ scattering 
amplitude. Assuming isospin symmetry, 
there is only one independent scattering amplitude $T(s,t,u)$ which can be 
identified as the $\pi^+(p_1)\pi^-(p_2)\rightarrow\pi^0(k_1)\pi^0(k_2)$ 
scattering amplitude, with the usual definitions of Mandelstam variables: 
$s=(p_1+p_2)^2$, $t=(p_1-k_1)^2$, $u=(p_1-k_2)^2$. All 
other amplitudes can then be expressed in terms of $T(s,t,u)$ and its 
crossings. For instance, the scattering amplitudes of definite isospin are
given by {\color{red}\cite{Petersen:1971ai}}
\begin{eqnarray}
T^{I=0}(s,t,u) \ali = \ali 3T(s,t,u)+T(t,s,u)+T(u,t,s) \, ,\nonumber \\
T^{I=1}(s,t,u) \ali = \ali T(t,s,u)-T(u,t,s)\, ,\nonumber \\
T^{I=2}(s,t,u) \ali = \ali T(t,s,u)+T(u,t,s)\, . \label{eq:TI}
\end{eqnarray}

\begin{figure}[h]
	\centering
	\includegraphics[scale=0.15]{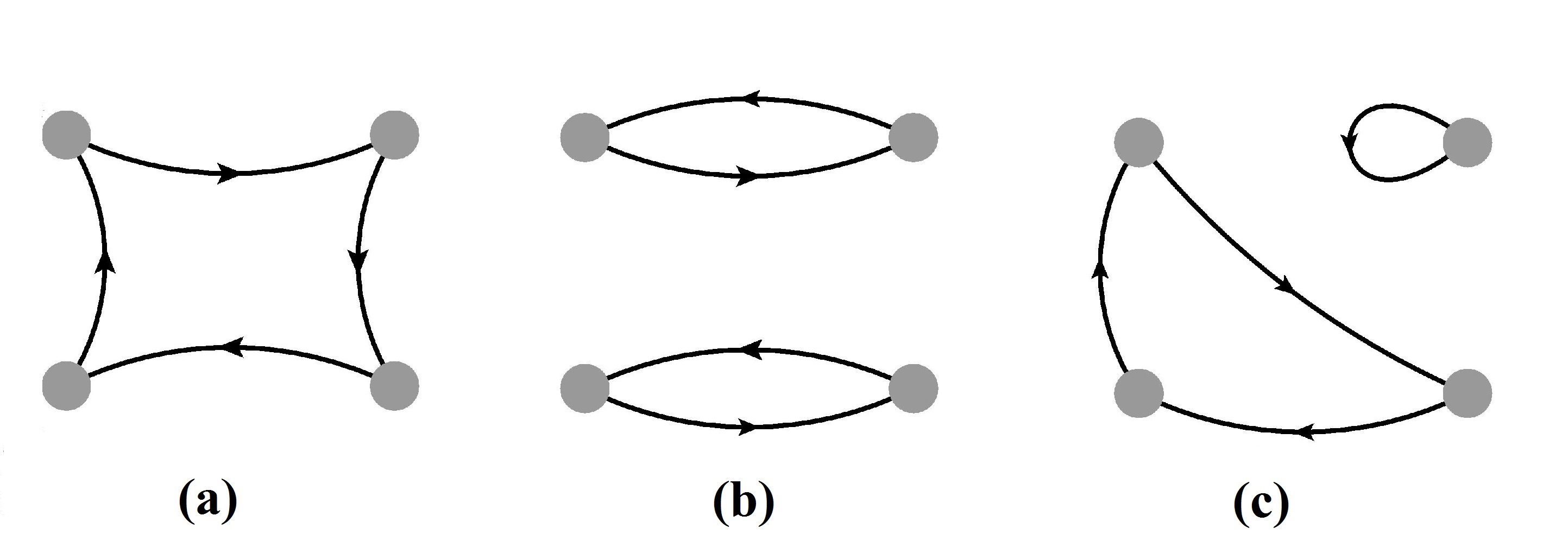}
	\caption{\label{fig:contraction}Three independent quark contraction 
diagrams in $\pi\pi$ scattering up to $\order{p^4}$. Each grey dot represents a 
$\bar{q}q$ state of definite flavor.}
\end{figure}

We can proceed to draw all possible Wick contraction diagrams that occur in 
$\pi\pi$ scattering. Realizing that each closed quark loop in the contraction 
diagram implies a flavor trace in the corresponding chiral Lagrangian, up to 
the $\order{p^4}$ level we shall retain diagrams that have at most two closed 
quark loops. There are only three independent contraction diagrams at this 
level, given in Fig. \ref{fig:contraction}; all other contraction diagrams can 
be obtained from these three by crossing. Furthermore, if we assume isospin 
symmetry, then the contribution from the third diagram vanishes. The reason is 
that SU$(N)$ generators are traceless so the contributions from all 
flavor-diagonal $\bar{q}q$  component to the upper right loop of the diagram 
sum up to zero. Therefore, the only two diagrams relevant in physical scattering 
processes are diagrams \ref{fig:contraction}(a) and \ref{fig:contraction}(b),
with the corresponding amplitudes labeled as $T_1(s,t,u)$ and $T_2(s,t,u)$,
respectively. For example,  the $\pi^+\pi^-\rightarrow\pi^0\pi^0$ 
amplitude can be written as
\begin{equation}
T(s,t,u)=T_1(s,t,u)+T_1(s,u,t)-T_1(u,t,s)+T_2(t,s,u)\, . \label{eq:Tdecompose}
\end{equation}
This can be seen by taking $\pi^+= u\bar{d}$, $\pi^-= d\bar{u}$, $\pi^0= (1/\sqrt{2})(u\bar{u}-d\bar{d})$ and 
performing all possible Wick contractions in the $\pi^+\pi^-\rightarrow\pi^0\pi^0$ amplitude.

It is not difficult to see that the simplest PQQCD that could make a clean 
separation of these two contraction diagrams will have four types of fermionic 
quarks due to the existence of four contraction lines. Meanwhile, two ghost 
quarks have to be included to keep the number of dynamical DOFs to be two, same 
as in the original SU(2) theory. Therefore, one concludes that the simplest 
PQQCD relevant to our work is SU$(4|2)$. The two contraction diagrams can be 
expressed in terms of scattering amplitudes in SU$(4|2)$ PQChPT as
\begin{eqnarray}
T_{1}(s,t,u) \ali = \ali 
T_{(u\bar{d})(d\bar{j})\rightarrow(u\bar{k})(k\bar{j})}(s,t,u)\, , \nonumber \\
T_{2}(s,t,u) \ali = \ali 
T_{(u\bar{d})(j\bar{k})\rightarrow(u\bar{d})(j\bar{k})}(s,t,u) \, .
\end{eqnarray}

\begin{figure}
\begin{centering}
	\includegraphics[scale=0.15]{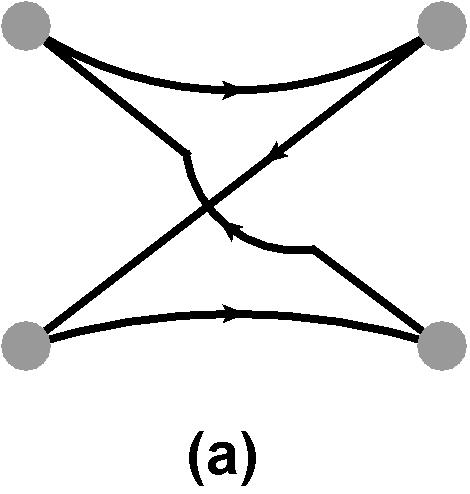}$\:\:\:$ 
\includegraphics[scale=0.15]{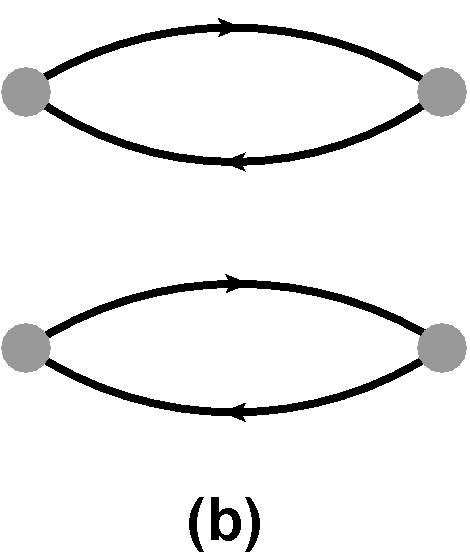}$\:\:\:$
	\includegraphics[scale=0.15]{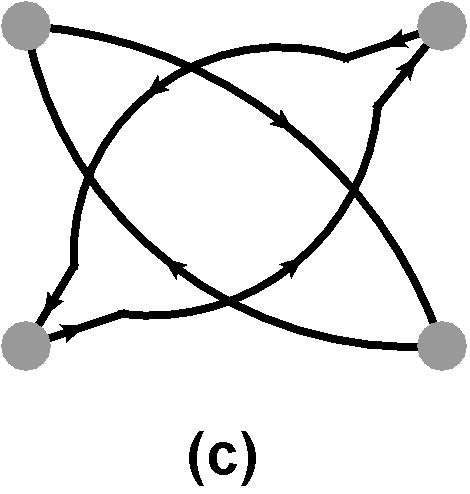}$\:\:\:$ 
\includegraphics[scale=0.15]{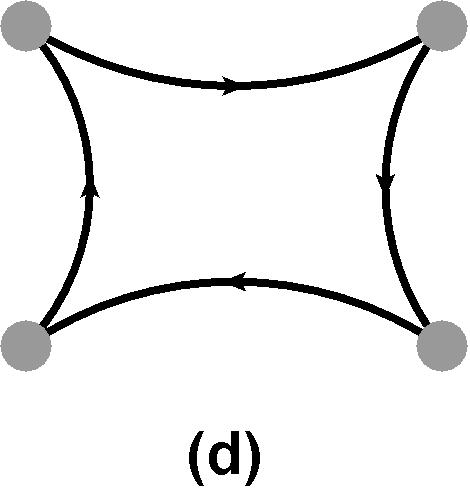}$\:\:\:$
	\includegraphics[scale=0.15]{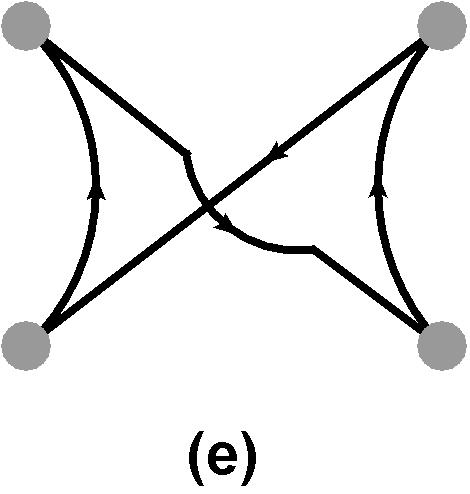}$\:\:\:$ 
\includegraphics[scale=0.15]{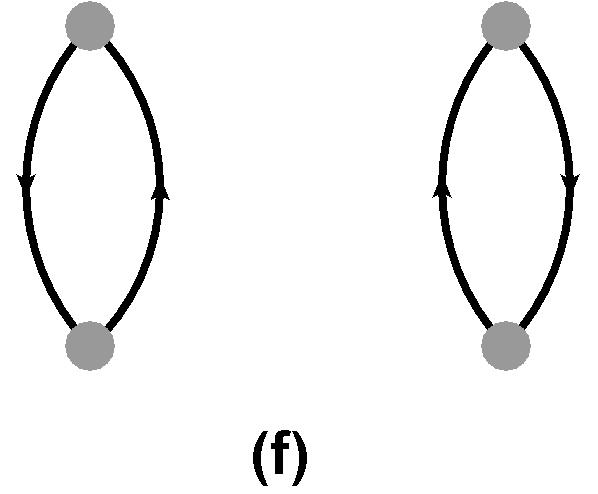}
	\par\end{centering}
\caption{\label{fig:connection}Contraction diagrams with definite type of 
contractions. Diagram (a) is crossed (C), diagrams (b) and (c) are direct (D),
diagrams (d) and (e) are rectangular (R) while diagram (f) is vacuum (V). 
They correspond to the amplitudes $T_1(u,t,s)$, $T_2(s,t,u)$, 
$T_2(s,u,t)$, $T_1(s,t,u)$, $T_1(s,u,t)$ and $T_2(t,s,u)$, respectively.}
\end{figure}

In connection with the lattice calculation, it is instructive to further 
classify $T_1(s,t,u),T_2(s,t,u)$ and their crossings into connected, 
singly-disconnected and doubly-disconnected diagrams, where connected,
singly-disconnected and doubly-disconnected diagrams have two, one and zero pair
of quark and anti-quark propagating from the initial to the final state,
respectively.
One sees that 
$T_1(u,t,s),\,T_2(s,t,u)$ and $ T_2(s,u,t)$ are connected diagrams, $T_1(s,t,u)$ 
and $T_1(s,u,t)$ are singly-disconnected diagrams while $T_2(t,s,u)$ is 
doubly-disconnected diagram. Adopting the notation of Ref. \cite{Kuramashi:1993ka}, 
we name $T_1(u,t,s)$ as crossed (C), $T_2(s,t,u)$ and $T_2(s,u,t)$ as direct 
(D), $T_1(s,t,u)$ and $T_1(s,u,t)$ as rectangular (R) and $T_2(t,s,u)$ as 
vacuum (V) diagrams, respectively (see Fig. \ref{fig:connection}). One can 
therefore decompose the isospin amplitudes according to the type of contraction 
of the diagrams as
\begin{eqnarray}
T^{I=0}(s,t,u)\ali =\ali \left[-T_1(u,t,s)\right]+\left[T_2(s,t,u)+T_2(s,u,t)\right]
+\left[3\left(T_1(s,t,u)+T_1(s,u,t)\right)\right]+\left[3T_2(t,s,u)\right]
\nonumber\\
&\equiv&T^{I=0}_C(s,t,u)\:+\qquad T^{I=0}_D(s,t,u)\qquad\:\:\:+\qquad 
T^{I=0}_R(s,t,u)\qquad\qquad\:+T^{I=0}_V(s,t,u)\, , \nonumber\\
T^{I=1}(s,t,u)\ali =\ali \left[T_2(s,t,u)-T_2(s,u,t)\right]+\left[2\left(T_1(s,t,
u)-T_1(s,u,t)\right)\right], \nonumber\\
&\equiv&\qquad T^{I=1}_D(s,t,u)\qquad\:\:\:+\qquad T^{I=1}_R(s,t,u)\,, 
\nonumber\\
T^{I=2}(s,t,u)\ali =\ali \left[2T_1(u,t,s)\right]+\left[T_2(s,t,u)+T_2(s,u,t)\right]
\nonumber\\
&\equiv&T^{I=2}_C(s,t,u)+\qquad T^{I=2}_D(s,t,u) \, ,
\end{eqnarray}
where we have defined the individual components $\{T^I_X(s,t,u)\}$ ($X=D,C,R,V$)
for each isospin amplitude $T^I(s,t,u)$. One observes that the $I=0$ amplitude
has all connected, singly-disconnected and doubly-disconnected pieces, the $I=1$ 
amplitude has connected and singly-disconnected pieces, while the $I=2$ 
amplitude has only the connected piece. It is interesting to observe that the
doubly-disconnected (V) contribution to the $I=0$ amplitude starts is
an NLO effect while the singly-disconnected (R) diagrams are at LO, as was
already noticed in Ref.~\cite{Guo:2013nja}.
Therefore, we expect that if the most difficult doubly-disconnected diagram was
neglected in a lattice calculation for $I=0$ 
$\pi\pi$ one would still get results in the correct ball park, while discarding 
the singly-disconnected diagram would lead to erroneous results. In the case of 
$I=1$ the singly-disconnected piece actually dominates over the connected piece 
because the former is of order $\order{p^2}$ while the latter is of order 
$\order{p^4}$ in chiral power counting.

\section{Analytical results}
In this section we present the main analytical results of this work. First we 
define all the special quantities that appear in the loop calculations.
The ultraviolet (UV) divergences and chiral logarithms are encoded
in\footnote{Throughout this paper, we take $M_\pi$ to be the charged pion mass 
and $F_\pi=92.2$ MeV.}:
\begin{eqnarray}
\lambda \ali\equiv \ali 
-\frac{1}{32\pi^{2}}(\frac{2}{4-d}+\ln4\pi-\gamma+1)\, ,\nonumber \\
\mu_{\pi} \ali \equiv \ali
-\frac{M_{\pi}^{2}}{32\pi^{2}F_{\pi}^{2}}\ln\frac{\mu^{2}}{M_{\pi}^{2}}\, .
\end{eqnarray}
We also need the Passarino--Veltman two-point function \cite{Passarino:1978jh}:
\begin{equation}
B_{0}(p^{2},m_{1}^{2},m_{2}^{2})\equiv\frac{(2\pi)^{4-d}}{i\pi^{2}}\int 
d^{d}k\frac{1}{(k^{2}-m_{1}^{2}+i\varepsilon)((k+p)^{2}-m_{2}^{2}+i\varepsilon)}
\, .
\end{equation}
Since the only mass that appears in this work is $M_{\pi}$,  we may
define the function $J_{\pi\pi}$ as
\begin{equation}
J_{\pi\pi}(p^{2})\equiv\frac{1}{16\pi^{2}}B_{0}(p^{2},M_{\pi}^{2},M_{\pi}^{2})\,.
\end{equation}
The finite piece in $J_{\pi\pi}(p^{2})$ can  be determined by explicitly 
isolating
its infinite piece proportional to $\lambda$
\begin{equation}
J_{\pi\pi}(p^{2})=-2\lambda+J_{\pi\pi}^{r}(p^{2}) \, .
\end{equation}

Next we shall quote the results for the $\beta$-functions of the $O(p^{4})$ 
LECs in SU$(4|2)$.
Most of them are identical to those in SU(2) because when SU$(4|2)$
is used to calculate processes involving only pions as external legs,
it should give an identical result with SU(2). The renormalized LECs
are defined through the bare LECs by
\begin{equation}
L_{i}^\mathrm{PQ}=L_{i}^{\mathrm{PQ},r}+\lambda\Gamma_{i}\, ,
\end{equation}
where the $\beta$-function coefficients $\Gamma_{i}$ in SU$(4|2)$ are
given in Table~\ref{tab:Coefficients-of-divergence}. One may also define the 
renormalized physical LECs $\{l_1^r,l_2^r,l_3^r,l_4^r,l_7^r\}$ through 
Eq.~\eqref{eq:physicalLEC} by replacing 
$L_i^\mathrm{PQ}\rightarrow L_i^{\mathrm{PQ},r}$. It is also customary to  
write the physical SU(2) LECs in terms of the scale-independent LECs
$\{\bar{l}_i\}$~\cite{Gasser:1983yg},
\begin{equation}
l_i^r=\gamma_i\left(\frac{\bar{l}_i}{32\pi^2}+\frac{F_\pi^2}{M_\pi^2}
\mu_\pi\right)
\end{equation}
for $i=1,...,4$, where $\gamma_1=1/3$, $\gamma_2=2/3$, $\gamma_3=-1/2$ and 
$\gamma_4=2$.

\begin{table}
	\begin{centering}
		\begin{tabular}{|c|c|c|c|c|c|c|c|c|c|}
			\hline 
			$i$ & 0 & 1 & 2 & 3 & 4 & 5 & 6 & 7 & 8\tabularnewline
			\hline 
			$\Gamma_{i}$ & $\frac{1}{24}$ & $\frac{1}{12}$ & 
$\frac{1}{6}$ & 0 & $\frac{1}{8}$ & $\frac{1}{4}$ & $\frac{3}{32}$ & 0 & 
0\tabularnewline
			\hline 
		\end{tabular}
		\par\end{centering}
	\caption{\label{tab:Coefficients-of-divergence}Coefficients of the  
	UV divergence in the SU$(4|2)$ PQChPT.}
	
\end{table}

As a first application and check of the theory, we compute the renormalization
of the wave function, mass and
decay constant of the pion in SU$(4|2)$. The outcomes are exactly the same as
in SU(2) as they should be and are given in {\color{red}\cite{Gasser:1983yg}}
\begin{eqnarray}
	Z_{\pi} \ali = \ali 
1+\frac{4}{3}\mu_{\pi}-\frac{2M_{\pi}^{2}}{F_{\pi}^{2}}l_4^r-\frac{8M_{\pi}^{2}}
{3F_{\pi}^{2}}\lambda \, ,\nonumber\\
	M_{\pi}^{2} \ali = \ali 
(M_{\pi}^{2})_{0}\left(1-\frac{M_\pi^2}{32\pi^2F_\pi^2}\bar{l}
_3\right), \nonumber\\
	F_{\pi} \ali = \ali 
F_{0}\left(1+\frac{M_\pi^2}{16\pi^2F_\pi^2}\bar{l}_4\right).
\end{eqnarray}
Not surprisingly, these expressions depend only on the physical LECs 
$\{l_i^r\}$. Further, although these quantities are calculated for pions, 
they apply equally to all Goldstone bosons made up 
of only sea or valence quarks, i.e. $\phi_1,\ldots,\phi_{15}$,  due 
to exact SU(4) symmetry.

The first two contraction diagrams in Fig.~\ref*{fig:contraction} can now be 
computed in SU$(4|2)$. They are
\begin{eqnarray}
T_{1}(s,t,u) \ali = \ali  
\frac{2M_{\pi}^{2}-u}{2F_{\pi}^{2}}+\left(\frac{3u-4M_{\pi}^{2}}{3F_{\pi}^{2}}
\right)\mu_{\pi}+\left(\frac{2M_{\pi}^{4}-4M_{\pi}^{2}(s+t)+s(2s+t)}{12F_{\pi}^{
4}}\right)J_{\pi\pi}^{r}(s)\nonumber \\
\ali \ali 
+\left(\frac{2M_{\pi}^{4}-4M_{\pi}^{2}(s+t)+t(2t+s)}{12F_{\pi}^{4}}\right)J_{
\pi\pi}^{r}(t)+\frac{4}{F_{\pi}^{4}}\left(s^{2}+t^{2}+u^{2}-4M_{\pi}^{4}
\right)L_{0}^{\mathrm{PQ},r}\nonumber \\
\ali \ali 
+\frac{2}{F_{\pi}^{4}}\left(4M_{\pi}^{2}u+s^{2}+t^{2}-8M_{\pi}^{4}\right)L_{3}^{
\mathrm{PQ},r}-\frac{4M_{\pi}^{2}u}{F_{\pi}^{4}}L_{5}^{\mathrm{PQ},r}+\frac{16M_
{\pi}^{4}}{F_{\pi}^{4}}L_{8}^{\mathrm{PQ},r}\nonumber \\
\ali \ali 
-\frac{M_{\pi}^{4}}{72\pi^{2}F_{\pi}^{4}}+\frac{M_{\pi}^{2}u}{96\pi^{2}F_{\pi}^{
4}}+\frac{2u^{2}-s^{2}-t^{2}}{576\pi^{2}F_{\pi}^{4}} \, ,\nonumber \\
T_{2}(s,t,u) \ali = \ali  
\frac{(s-2M_\pi)^2}{4F_{\pi}^{4}}J_{\pi\pi}^{r}(s)+\frac{(u-2M_\pi)^2}{4F_{\pi}^
{4}}J_{\pi\pi}^{r}(u)+\left(\frac{2M_{\pi}^{4}+t^{2}}{4F_{\pi}^{4}}\right)J_{
\pi\pi}^{r}(t)\nonumber \\
\ali \ali 
+\frac{4}{F_{\pi}^{4}}\left(4M_{\pi}^{4}-s^{2}-t^{2}-u^{2}\right)L_{0}^{\mathrm{
PQ},r}+\frac{2}{F_{\pi}^{4}}\left(t-2M_\pi^2\right)^2(l_1^r-2L_3^{\mathrm{PQ},r}
)\nonumber \\
\ali \ali 
+\frac{1}{F_{\pi}^{4}}\left(s^{2}+u^{2}+4M_{\pi}^{2}t-8M_{\pi}^{4}
\right)l_2^r+\frac{2M_{\pi}^{2}}{F_{\pi}^{4}}\left(t-2M_{\pi}^{2}
\right)(l_4^r-4L_5^{\mathrm{PQ},r})\nonumber\\
\ali\ali +\frac{2M_{\pi}^{4}}{F_{\pi}^{4}}(l_3^r+l_4^r-8L_8^{\mathrm{PQ},r}) \,
.
\label{eq:T1T2}
\end{eqnarray}
Each amplitude is finite and scale-independent. One sees that $T_{1}(s,t,u)$ 
begins at $O(p^{2})$ while $T_{2}(s,t,u)$ begins at $O(p^{4})$ as we expected 
from counting the number of closed quark loops. Moreover, $T_{1}(s,t,u)$ is 
symmetric with respect to $s\leftrightarrow t$ while $T_{2}(s,t,u)$ is 
symmetric with respect to $s\leftrightarrow u$. One also observes that both 
amplitudes have a dependence on the unphysical LECs, reflecting the 
fact that the separation of the scattering amplitude into different contraction 
diagrams is in fact an artificial process. The outcome of this separation is, however, unique because 
each contraction can be represented by a scattering amplitude in the PQChPT.

As a cross-check, we also computed the 
$\pi^{+}\pi^{-}\rightarrow\pi^{0}\pi^{0}$ scattering amplitude to $\order{p^4}$ 
using SU(2) ChPT and confirmed the correctness of Eq.~\eqref{eq:Tdecompose}. 
As expected, the dependence on the unphysical LECs drops out as far as the 
scattering amplitudes of pions are concerned. We also cross-checked
Eq.~\eqref{eq:T1T2} using the formulation briefly discussed in \ref{sec:phi0}.

It will be instructive to display numerical results showing the relative 
importance of different contractions that can be directly contrasted to the 
lattice results. The main problem, however, is that what is usually displayed in 
lattice plots are correlation functions $C(t,t')=\left\langle 
\mathcal{O}(t)\mathcal{O}'^\dagger(t')\right\rangle$ of the interpolating field 
operators $\mathcal{O}$ and $\mathcal{O}'$. Such operators are not the 
$\pi\pi$ field operators in the traditional sense although they have the same 
quantum numbers, not to say different lattice studies may choose different 
forms of interpolating operators of their convenience. There is hence no 
universal matching rule between $C(t,t')$ and physical scattering amplitude (on 
the other hand, carefully constructed ratios between correlation functions may 
be directly mapped to amplitudes in field theory). With these considerations, 
we decide to display the partial wave of the isospin amplitudes with a definite 
type of contraction:
\begin{equation}
T^{IJ}_{X}(s)\equiv 
\frac{1}{64\pi}\int_{-1}^{+1}T^I_X\left(s,t(s,\cos\theta),u(s,
\cos\theta)\right)P_J(\cos\theta)d\cos\theta
\end{equation}
where $X$=D,C,R,V denotes the type of contractions while $P_J(x)$ is the 
Legendre polynomial. They have obvious physical meanings and can be easily 
reconstructed from lattice data. Furthermore, the definition of partial waves 
does not require the unitarity of the amplitude so it is well-defined for each 
contraction separately. We shall display both the energy dependence of the 
partial waves as well as the scattering lengths $a_X^{IJ}$ which are given by
\begin{equation}
a_X^{IJ}=\lim_{q^2\rightarrow 
0}\frac{\mathrm{Re}T_X^{IJ}(4M_\pi^2+4q^2)}{(q^2)^J},
\label{eq:ax}
\end{equation}
where we have expressed the Mandelstam $s$ around the pion threshold 
as $s=4(M_\pi^2+q^2)$.

Using Eqs.~\eqref{eq:TI} and \eqref{eq:T1T2}, one is able to compute the 
contribution to the scattering lengths from each contraction $X$, and the total 
(i.e. physical) scattering length $a^{IJ}_{\mathrm{tot}}=\sum_X a^{IJ}_X$. The 
full results are given in \ref{sec:scatteringlength}, and one may readily check 
the scale independence of each component. The full expressions for 
$a^{IJ}_{\mathrm{tot}}$ agree with Ref. \cite{Gasser:1983yg} as long as 
everything is expressed in terms of physical quantities. All such expressions 
above are, in principle, concrete predictions of PQChPT that can be directly 
contrasted with the lattice results. However, the degree of precision for each 
$a^{IJ}_X$ varies depending on the the LECs involved. It is therefore desirable 
to construct combinations that are least affected by the uncertainties of the 
LECs. In particular, one may wish to avoid any dependence on the $\bar{l}_3$, 
$L^{\mathrm{PQ,r}}_0$ and $L^{\mathrm{PQ,r}}_3$ that suffer the most 
uncertainties in data/lattice fitting (see the discussion in the next section). 
With such considerations, we find the following combinations (apart from the 
total scattering length $a^{IJ}_{\mathrm{tot}}$) particularly interesting:
\begin{eqnarray}
a^{00}_V-\frac{3}{2}a^{00}_D\ali =\ali \frac{M_\pi^4}{\pi 
F_\pi^4}\left[\frac{3\bar{l}_4}{64\pi^2}-3L^{\mathrm{PQ,r}}_5+\frac{9}{512\pi^2}
+\frac{3F_\pi^2}{4M_\pi^2}\mu_\pi\right],\nonumber\\
a^{00}_R+6a^{00}_C-\frac{3M_\pi^2}{8\pi F_\pi^2}\ali =\ali \frac{M_\pi^4}{\pi 
F_\pi^4}\left[3L^{\mathrm{PQ,r}}_5+\frac{3}{256\pi^2}-\frac{3F_\pi^2}{4M_\pi^2}
\mu_\pi\right],\nonumber\\
M_\pi^2a^{11}_R+\frac{8}{3}a^{00}_C-30M_\pi^4a^{02}_C-\frac{M_\pi^2}{8\pi 
F_\pi^2}\ali =\ali \frac{M_\pi^4}{\pi 
F_\pi^4}\left[\frac{5L^{\mathrm{PQ,r}}_5}{3}-\frac{4L^{\mathrm{PQ,r}}_8}{3}
-\frac{479}{34560\pi^2}-\frac{5F_\pi^2}{12M_\pi^2}\mu_\pi\right],\nonumber\\
M_\pi^4a^{02}_R-3M_\pi^4a^{02}_C+\frac{4}{5}a^{00}_C-\frac{M_\pi^2}{40\pi 
F_\pi^2}\ali =\ali \frac{M_\pi^4}{\pi 
F_\pi^4}\left[\frac{2L^{\mathrm{PQ,r}}_5}{5}-\frac{2L^{\mathrm{PQ,r}}_8}{5}
-\frac{31}{9600\pi^2}-\frac{F_\pi^2}{10M_\pi^2}\mu_\pi\right],\nonumber\\
M_\pi^4a^{02}_V+6M_\pi^4a^{02}_C-\frac{4}{5}a^{00}_C+\frac{M_\pi^2}{40\pi 
F_\pi^2}\ali =\ali \frac{M_\pi^4}{\pi 
F_\pi^4}\left[\frac{\bar{l}_2}{480\pi^2}-\frac{2L^{\mathrm{PQ,r}}_5}{5}+\frac{
2L^{\mathrm{PQ,r}}_8}{5}+\frac{1}{2400\pi^2}+\frac{F_\pi^2}{10M_\pi^2}
\mu_\pi\right].
\label{eq:combination}
\end{eqnarray}
We choose to subtract out all the $\order{p^2}$ contributions in the 
construction of these combinations above so that none of the contraction 
diagrams is suppressed by trivial power counting. These relations are useful 
because they relate the singly- and doubly-disconnected contributions ($a_R$ 
and $a_V$) to the connected contributions ($a_D$ and $a_C$) where the latter 
can be determined most accurately, and furthermore the uncertainties due to LECs
are minimal. They can be used to check the accuracy of lattice studies on the 
disconnected pieces. 

Before ending this section, we shall also present the analytic expressions for
the imaginary part of each partial wave in the physical region, i.e.
$s\geq4M_\pi^2$, as they are LEC-free and are examples of clean PQChPT
predictions at one loop. As these are leading order results, they
are known to not be very precise.
By direct inspection one finds that the only partial
waves with a nonvanishing imaginary part are:
\begin{eqnarray}
\mathrm{Im}T^{00}_D(s)\ali =\ali \frac{(s-2M_\pi^2)^2}{64\pi 
F_\pi^4}\mathrm{Im}J^r_{\pi\pi}(s) \, ,\nonumber\\
\mathrm{Im}T^{00}_R(s)\ali =\ali \frac{3(s^2-4M_\pi^4)}{128\pi 
F_\pi^4}\mathrm{Im}J^r_{\pi\pi}(s) \, ,\nonumber\\
\mathrm{Im}T^{00}_V(s)\ali =\ali \frac{3(s^2+2M_\pi^4)}{128\pi 
F_\pi^4}\mathrm{Im}J^r_{\pi\pi}(s) \, ,\nonumber\\
\mathrm{Im}T^{11}_R(s)\ali =\ali \frac{(s-4M_\pi^2)^2}{576\pi 
F_\pi^4}\mathrm{Im}J^r_{\pi\pi}(s) \, ,\nonumber\\
\mathrm{Im}T^{20}_D(s)\ali =\ali \frac{(s-2M_\pi^2)^2}{64\pi 
F_\pi^4}\mathrm{Im}J^r_{\pi\pi}(s) \, ,
\end{eqnarray}
where 
\begin{equation}
\mathrm{Im}J^r_{\pi\pi}(s)=\Theta(s-4M_\pi^2)\frac{1}{16\pi}\sqrt{\frac{
s-4M_\pi^2}{s}}\,.
\end{equation}
It is also worthwhile to mention that the physical $\pi\pi$ partial waves with 
definite isospin satisfy the single-channel perturbative unitarity 
relation:
 \begin{equation}
  \mathrm{Im}T_\mathrm{NLO}^{IJ}(s)=\frac{2|\vec{p}\,|}{\sqrt{s}}
  \left|T_\mathrm{LO}^{IJ}(s)\right|^2 ,
 \end{equation}
where $|\vec{p}|$ is the center-of-mass momentum, but the components 
from a specific contraction in general do not satisfy such a relation because 
they do not have definite isospin. Instead, since they can be written as 
scattering amplitudes in PQChPT, they should satisfy a general 
multi-channel perturbative unitarity relation:
\begin{equation}
  \mathrm{Im}T^J_{\mathrm{NLO},ab\rightarrow cd}(s)=\sum_{e,f}\alpha_{ef}
  \frac{2|\vec{p}\,|}{\sqrt{s}}T^J_{\mathrm{LO},ef\rightarrow cd}(s)
  T^{J*}_{\mathrm{LO},ab\rightarrow ef}(s),
\end{equation}
where $\alpha_{ef}=1(2)$ if $e$ and $f$ are (are not) identical particles.	

\section{Numerical Results}

\begin{table}
	\begin{centering}
		\begin{tabular}{|c|c|}
			\hline 
			\hline 
			$\bar{l}_1$ &  $-0.4(6)$ \\
			$\bar{l}_2$ & 4.3(1)\\
			$\bar{l}_3$ & 3.0(8)\\
			$\bar{l}_4$ & 4.4(2)\\
			$10^3L_{0}^{\mathrm{PQ},r}$ & 1.0(1.1)\\
			$10^3(L_{3}^{\mathrm{PQ},r}+2L_{0}^{\mathrm{PQ},r})$ & $-1.56(87)$\\
			$10^3L_{5}^{\mathrm{PQ},r}$ & 0.501(43)\\
			$10^3L_{8}^{\mathrm{PQ},r}$ & 0.581(22)\\		
			\hline 
			\hline 
		\end{tabular}
		\par\end{centering}
	\caption{\label{tab:LECs}The LECs relevant to our work. Values of 
$\{\bar{l}_i\}$ are taken from Refs.~\cite{Colangelo:2001df,Bijnens:2014lea}. 
$\{L^{\mathrm{PQ,r}}_5,L^{\mathrm{PQ,r}}_8\}$ and 
$\{L^{\mathrm{PQ,r}}_0,L^{\mathrm{PQ,r}}_3\}$ are obtained from the NLO and NNLO 
fits to lattice data, respectively, in Ref.~\cite{Boyle:2015exm} at $\mu=1$~GeV
(see the discussion in the text).}	
\end{table}

In this section we present our numerical results. For this purpose we need
the values for both the physical LECs $\bar{l}_i$ as well as the unphysical ones
$\{L_i^{\mathrm{PQ},r}\}$ ($i=0,3,5,8$). $\bar{l}_1$, $\bar{l}_2$ and
$\bar{l}_4$ were obtained by sophisticated analysis of the $\pi\pi$ scattering
data~\cite{Colangelo:2001df}. $\bar{l}_3$ has a very large uncertainty that is,
to some extent, reduced by the inclusion of lattice
results~\cite{Bijnens:2014lea}. For the physical LECs, we quote the values from
Ref.~\cite{Bijnens:2014lea} that combines fits to data and lattice results,
expressed in terms of the scale-independent LECs. For the unphysical LECs, we
quote the recent lattice results by the RBC-UKQCD Collaboration from $N_f=2+1$
domain wall QCD~\cite{Boyle:2015exm}, fitted to the pion mass and decay constant
calculated at NLO and next-to-next-to-leading order (NNLO) PQChPT including
the NLO finite volume corrections.
The underlying principle is that one is free to vary the sea and valence quark
masses independently on the lattice, and the outcome of such a manipulation can be
matched to PQChPT with non-degenerate Goldstone particles. For
$L^{\mathrm{PQ,r}}_5$ and $L^{\mathrm{PQ,r}}_8$, we take their values from the
NLO fit that was performed with satisfactory precision. On the other hand,
$L^{\mathrm{PQ,r}}_0$ and $L^{\mathrm{PQ,r}}_3$ can only be obtained from an
NNLO fit (because they do not appear in the NLO mass and decay constant
correction) so that their precisions from fitting are less satisfactory. We
choose the renormalization scale at $\mu=1$~GeV and pick the data with a 450~MeV
mass cut on the unitary pion because the resuls with this cut seem to reproduce the
experimental values of $\bar{l}_i$ a bit better. We summarize our numerical
choice of the LECs in Table~\ref{tab:LECs}.\footnote{The readers should be aware
of the difference in the definition of
$\{L_1^{\mathrm{PQ}},L_2^{\mathrm{PQ}},L_3^{\mathrm{PQ}}\}$ between
Ref.~\cite{Boyle:2015exm} and ours. We define them in such a way that the
$L_0^{\mathrm{PQ}}$-dependence vanishes in SU(2).}
\begin{table}
	\begin{centering}
		\begin{tabular}{|c|c|c|c|c|c|}
			\hline 
			& $10^2a_{X}^{00}$ & $10^2a_{X}^{20}$ & 
$10^2M_{\pi}^{2}a_{X}^{11}$ & $10^4M_{\pi}^{4}a_{X}^{02}$ & 
$10^4M_{\pi}^{4}a_{X}^{22}$\tabularnewline
			\hline 
			\hline 
			D & $0.35\pm0.24$ & $0.35\pm0.24$ & $0.02\pm0.26$ & 
$3.5\pm2.0$& $3.5\pm2.0$\tabularnewline
			\hline 
			C & $2.41\pm0.12$ & $-4.81\pm0.23$ & 0 & $0.95\pm0.96$& 
$-1.9\pm1.9$\tabularnewline
			\hline 
			R & $14.8\pm0.7$ & 0 & $3.59\pm0.26$ & 
$6.7\pm7.8$&0\tabularnewline
			\hline 
			V & $2.48\pm0.38$ & 0 & 0 & $0.8\pm7.3$&0\tabularnewline
			\hline 
			Total & $20.0\pm0.2$ & $-4.46\pm0.07$ & $3.61\pm0.04$ & 
$11.9\pm0.8$& $1.54\pm0.71$\tabularnewline
			\hline 
		\end{tabular}
		\par\end{centering}
	\caption{\label{tab:aIJX}PQChPT predictions of the scattering lengths 
from each contraction.}
\end{table}

The PQChPT predictions of the scattering lengths $a^{IJ}_X$ are given in
Table~\ref{tab:aIJX}. We include only the uncertainties of the LECs from
Table~\ref{tab:LECs} in our error analysis and compute the final uncertainties
using a simple error propagation formula where the errors of the LECs are 
assumed to be uncorrelated and are combined in a quadrature. One sees that the
uncertainty of each entry is quite large due to the existence of the
badly determined LECs $\bar{l}_3$, $L^{\mathrm{PQ,r}}_0$ and
$L^{\mathrm{PQ,r}}_3$. On the other hand, the precision of the combinations
given in Eq.~\eqref{eq:combination} is much higher:
\begin{eqnarray}
a^{00}_V-\frac{3}{2}a^{00}_D\ali =\ali (1.96\pm0.16)\times 10^{-2}\nonumber\\
a^{00}_R+6a^{00}_C-\frac{3M_\pi^2}{8\pi F_\pi^2}\ali =\ali (2.00\pm0.02)\times 
10^{-2} \,,\nonumber\\
M_\pi^2a^{11}_R+\frac{8}{3}a^{00}_C-30M_\pi^4a^{02}_C-\frac{M_\pi^2}{8\pi 
F_\pi^2}\ali =\ali (6.38\pm0.13)\times 10^{-3} \,,\nonumber\\
M_\pi^4a^{02}_R-3M_\pi^4a^{02}_C+\frac{4}{5}a^{00}_C-\frac{M_\pi^2}{40\pi 
F_\pi^2}\ali =\ali (1.47\pm0.03)\times 10^{-3} \,,\nonumber\\
M_\pi^4a^{02}_V+6M_\pi^4a^{02}_C-\frac{4}{5}a^{00}_C+\frac{M_\pi^2}{40\pi 
F_\pi^2}\ali =\ali (-4.39\pm0.47)\times 10^{-4} \,.\label{eq:numericcomb}
\end{eqnarray}
One observes that their absolute uncertainties are much smaller than those of
$a_V^{00}$, $a_R^{00}$, $M_\pi^2a_R^{11}$, $M_\pi^4a_R^{02}$  and
$M_\pi^4a_V^{02}$ themselves, respectively. We would like to stress that this in
fact a nice example of the mutually beneficial interplay between EFT-based
theoretical studies and lattice QCD. On the one hand, Eq.~\eqref{eq:numericcomb}
serves as a useful check on the degree of accuracy of lattice studies in the
handling of the singly- and doubly-disconnected contributions. On the other
hand, since the connected diagrams can be calculated easily, they can 
be used to perform
NLO fits for the poorly-known unphysical LECs $L^{\mathrm{PQ,r}}_0$ and
$L^{\mathrm{PQ,r}}_3$ through $a^{IJ}_D$ and $a^{IJ}_C$ with the formulas given
in \ref{sec:scatteringlength}, which cannot be done by the fitting to the pion mass 
and decay constant at NLO. This may greatly reduce the uncertainties in such 
LECs, which will provide a more precise calculation of quantities in PQChPT
at $\order{p^4}$ including the disconnected contributions to  $\pi\pi$
scattering.

\begin{figure}[tbh]
	\begin{centering}
		\includegraphics[width=0.49\textwidth]{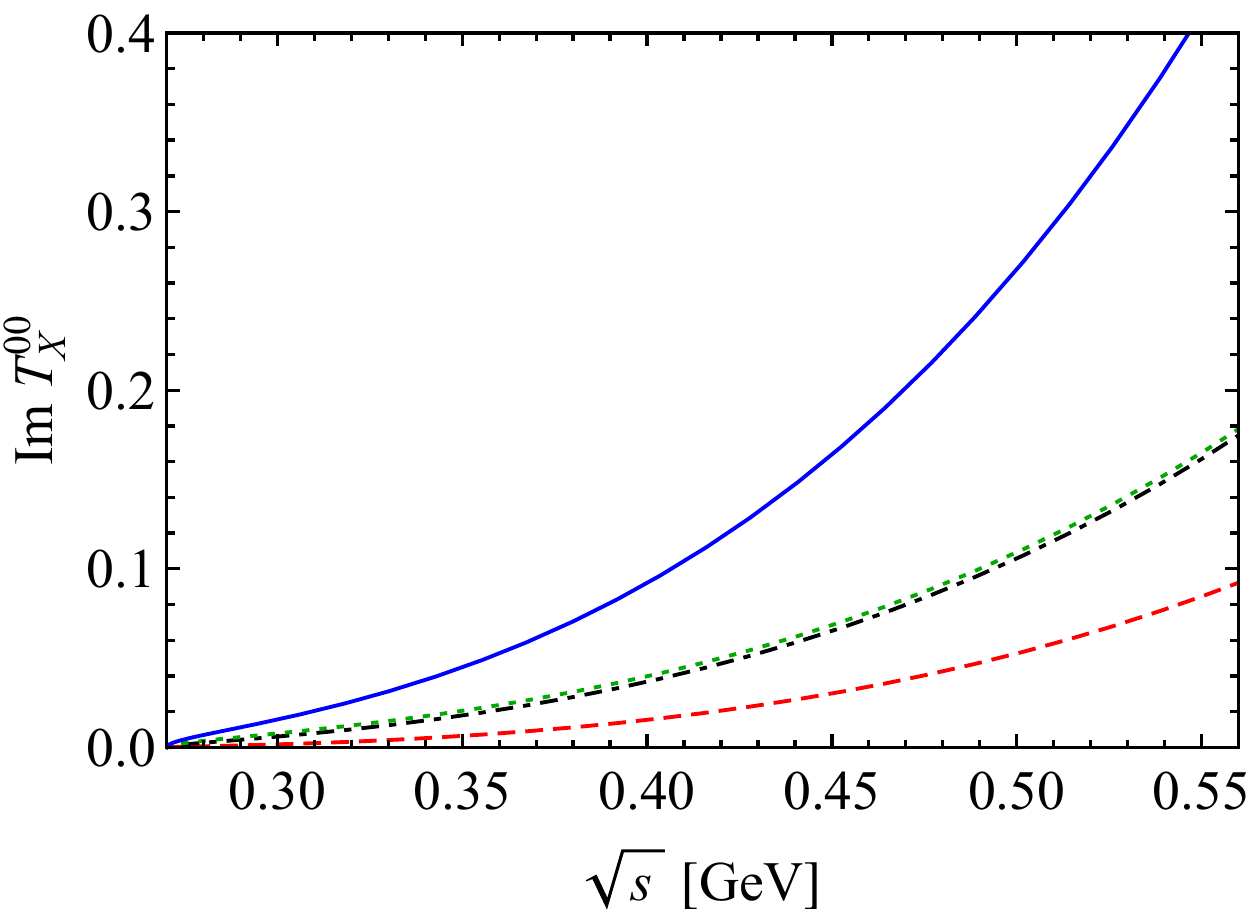}\hfill
		\includegraphics[width=0.5\textwidth]{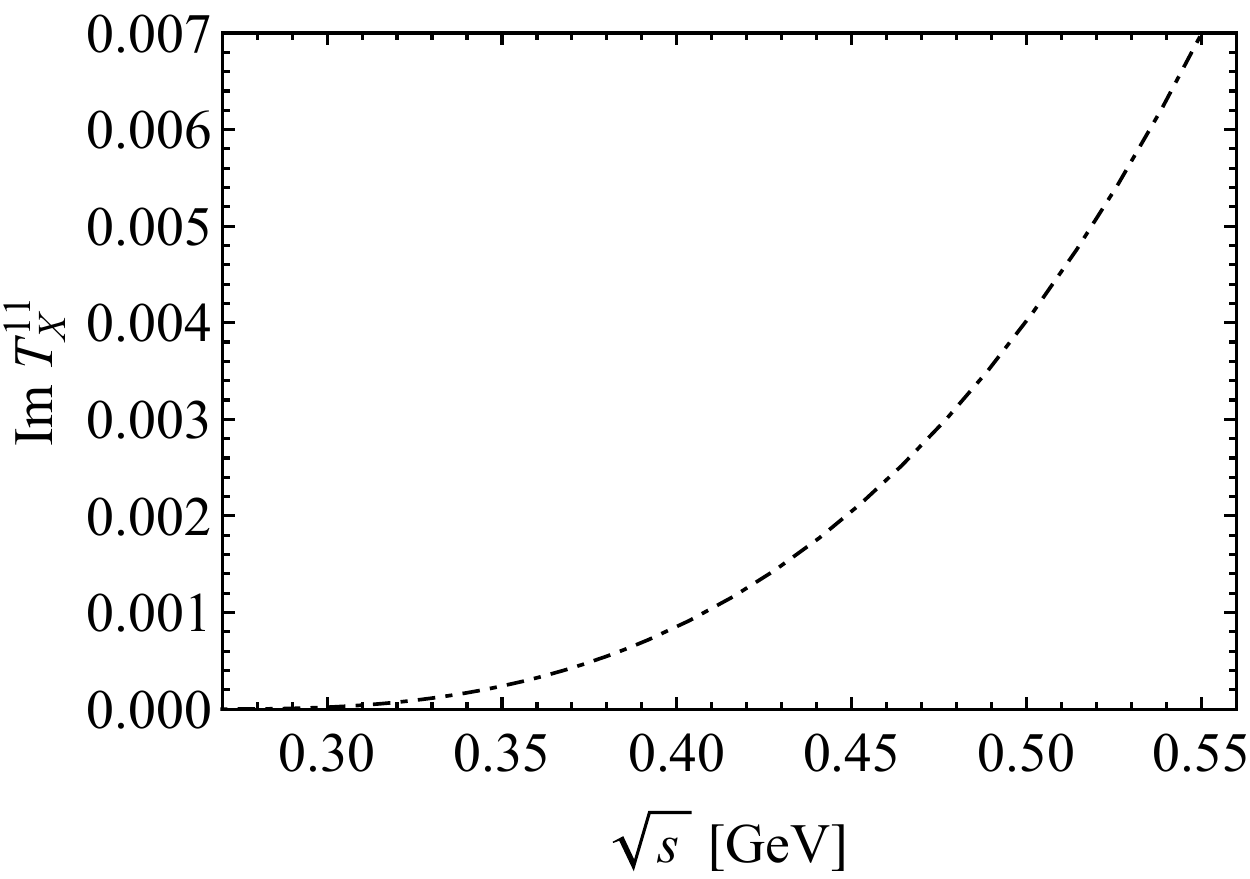}\\[2mm]
		\includegraphics[width=0.49\textwidth]{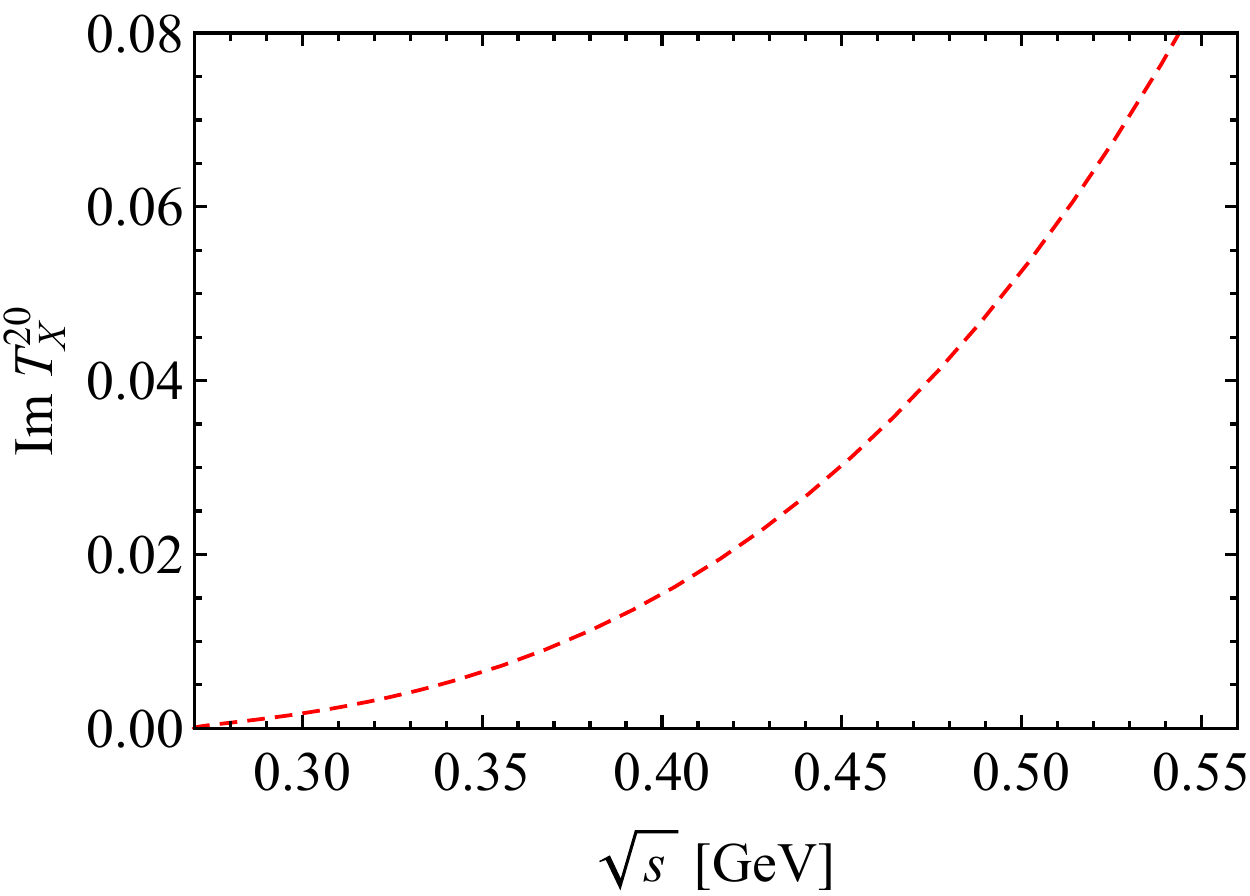}
		\par\end{centering}
	\caption{(Color online) Imaginary part of the direct (red dashed), 
rectangular (black dash-dotted), vacuum (green dotted) and total (blue solid, 
for $T^{00}$ only) contribution to the partial wave 
amplitude.\label{fig:imaginary}}
	\end{figure}

It will be also instructive to study, instead of just the threshold parameters,
the full $s$-dependence of the partial wave amplitudes. Unfortunately, for this
case there is no simple combination  that allows us to get rid of the
uncertainties brought up by $\bar{l}_3$, $L^{\mathrm{PQ,r}}_0$ and
$L^{\mathrm{PQ,r}}_3$ that in general grow with increasing $s$. Therefore, we
choose to present only the imaginary part of the partial waves because they are
LEC-free, as described in the previous section. The plots are given in
Fig.~\ref{fig:imaginary}. One observes that the imaginary part of the $I=0$
amplitude in the physical region comes from the direct, rectangular and vacuum
diagrams, while the imaginary part of $I=1(2)$ amplitude in the physical region
only originates from rectangular (direct) diagrams.

Before ending this section, we would like to point out that if the 
scattering lengths for the topologically decomposed contractions, defined in 
Eq.~\eqref{eq:ax}, can be extracted from lattice directly,\footnote{The 
theoretical framework remains to be worked out~\cite{progress}.} this could 
hint at a new direction to combine the theoretical and simulation efforts to 
find a more economical way to calculate the quantities involving disconnected 
contractions. Assuming that this is feasible, one may extract the
the badly known unphysical LECs $L_0^{\text{PQ},r}$ 
and $L_3^{\text{PQ},r}$ (see Table~\ref{tab:LECs})\footnote{It is 
worthwhile to mention that they always appear in the same linear
combination in all contraction-specific scattering lengths in a given 
$(I,J)$ channel.} from calculating the easier $D$- and $C$-type contractions, 
and feed them back into the 
formulae collected in~\ref{sec:scatteringlength}, or 
Eqs.~\eqref{eq:combination} and \eqref{eq:numericcomb}, to get the more 
difficult contractions.

\section{Summary}  

Isoscalar pion-pion scattering has been and still is a formidable challenge in lattice
calculations mainly due to the existence of doubly-disconnected/vacuum
Wick contractions which are extremely noisy.
In this work we have studied the contributions from different types of quark
contractions to the pion-pion scattering amplitudes by  means of SU$(4|2)$
PQChPT up to $\order{p^4}$. By suitably
choosing the quark content of the incoming and outgoing Goldstone particles, one
is able to probe the scattering amplitudes with any given contraction. We
express the direct, crossed, rectangular and vacuum contraction diagrams in
terms of two independent amplitudes $T_1(s,t,u)$, $T_2(s,t,u)$ and their
crossings. In order  to provide definite predictions, we derived the analytic
expressions for $J=0,1,2$ scattering lengths resulting from definite
contractions. On top of that, we constructed observables that are minimally
affected by the uncertainties of the low-energy constants in PQChPT, including
specific combinations of scattering lengths and the imaginary part of partial
wave amplitudes.

Our work is a nice example of a mutually-beneficial interplay between effective
field theory and lattice QCD. The ability of lattice QCD to isolate diagrams of
different contractions enables NLO fitting of certain unphysical LECs in PQChPT,
and hence greatly improves its predictive power. In particular, we have given
expressions in terms of these LECs for various scattering lengths decomposed
into different types of contractions. They can be used to extract the values of
the unphysical LECs in PQChPT. It would also be great if lattice calculations
could decompose the results of scattering lengths into different contractions so
that the results from PQChPT and lattice QCD can be cross-checked. In this
sense, our prediction may even serve as a measure of the accuracy level of
lattice calculation of quark disconnected diagrams.

There are lots of other hadronic quantities whose calculation in lattice
involves the expensive and noisy disconnected diagrams, such as the $\pi K$ and
$K\bar K$ scattering processes, the parity-odd pion-nucleon coupling and so on.
The same kind of interplay could be useful in the study of them. The necessary
PQChPT analysis will be done in future works.

\vspace{0.5cm}

\noindent {\bf Acknowledgements} 

We are grateful to Xu Feng and Liuming Liu for helpful discussions 
and to Xu Feng for a careful reading and valuable comments on the 
first version of this manuscript. We also thank the referee for
very useful comments.
FKG acknowledges the warm hospitality of the INPAC,  CYS acknowledges the
warm hospitality of the ITP of CAS, and UGM acknowledges warm hospitality
of the IHEP of CAS and ITP of CAS,  where part of this work was done.
This work is supported in part by NSFC and DFG through funds provided to the
Sino-German CRC 110 ``Symmetries and the Emergence of Structure in QCD" (NSFC
Grant No.~11621131001, DFG Grant No.~TRR110), by NSFC (Grant No.~11575110,
No.~11175115 and No.~11647601), by the Thousand Talents Plan for Young
Professionals, by the CAS Key Research Program of Frontier Sciences (Grant
No.~QYZDB-SSW-SYS013), by the CAS President's International Fellowship
Initiative (PIFI) (Grant No.~2017VMA0025), and by the Natural Science Foundation of
Shanghai under Grant  No. 15DZ2272100 and No. 15ZR1423100.

\begin{appendix}

\section{PQChPT by integrating out $\Phi_0$}
\label{sec:phi0}

For practical purposes, sometimes it is convenient to use a method different from
the one used in the main text to implement the constraint of
$\mathrm{Str}(\Phi)=0$. This can be achieved by adding to the LO Lagrangian in
Eq.~\eqref{eq:L2} a mass term for the singlet field $\Phi_0\equiv
\mathrm{Str}(\Phi)/\sqrt{2}$, $-m_0^2\Phi_0^2$~\cite{Sharpe:2001fh}, which is
later on integrated out by setting $m_0$ to infinity. In this formalism, the
matrix $U$ is written as $U=\exp{\left(i\sqrt{2} \Phi/F_0\right) }$, with 
\begin{equation}
  \Phi = \begin{pmatrix}
    \phi & \chi^\dag \\
    \chi & \tilde{\phi}
  \end{pmatrix} ,  \quad
  \phi = \begin{pmatrix}
    \eta_u & \pi^+ & \phi_{u\bar j} & \phi_{u\bar k} \\
    \pi^- & \eta_d & \phi_{d\bar j} & \phi_{d\bar k} \\
    \phi_{j\bar u} & \phi_{j\bar d} & \eta_j & \phi_{j\bar k} \\
    \phi_{k\bar u} & \phi_{k\bar d} & \phi_{k\bar j} & \eta_{k}
  \end{pmatrix}, \quad
  \tilde{\phi} = \begin{pmatrix}
   \eta_{\tilde{j}} & \phi_{\tilde{j} \bar{\tilde{k}}} \\
   \phi_{\tilde{k} \bar{\tilde{j}}} & \eta_{\tilde{k} } 
  \end{pmatrix}, \quad
  \chi = \begin{pmatrix}
  \phi_{\tilde{j} \bar u} & \phi_{\tilde{j} \bar d} & \phi_{\tilde{j} \bar j} &
  \phi_{\tilde{j} \bar k} \\
  \phi_{\tilde{k} \bar u} & \phi_{\tilde{k} \bar d} & \phi_{\tilde{k} \bar j} &
  \phi_{\tilde{k} \bar k}
  \end{pmatrix},
\end{equation}
where the diagonal elements $\eta_q$'s are neutral states made of $q\bar q$.
After integrating our the singlet component $\Phi_0$, one gets the propagators
of these neutral states. For the case of  SU$(4|2)$ with all quarks and ghost
quarks degenerate, they are rather simple:
\begin{equation}
  G^{ab}(k^2) = \frac{i}{k^2-(M_\pi^2)_0+i\varepsilon} \left( \delta^{ab} 
\epsilon^a -\frac12 \right) ,
\end{equation}
where $\epsilon^a=+1 (-1)$ for neutral particles that are made out of fermionic (bosonic) 
quark-antiquark pairs. The propagators for all other mesons take the normal form except for an
additional factor of $-1$ for states made of a pair of ghost quark and
anti-quark.

\section{\label{sec:scatteringlength}Scattering Lengths}

In this appendix we present the analytic expressions for the $J=0,1,2$ 
scattering lengths resulting from each contraction $X$ as well as the physical 
scattering length $a^{IJ}_{\mathrm{tot}}$. 

\subsection*{$I=0,J=0$:}
\begin{eqnarray}
a^{00}_D\ali =\ali -\frac{M_\pi^2}{8\pi 
F_\pi^2}\left[\mu_\pi+\frac{M_\pi^2}{F_\pi^2}\left(-\frac{\bar{l}_1}{24\pi^2}
-\frac{\bar{l}_2}{12\pi^2}+\frac{\bar{l}_3}{64\pi^2}+\frac{\bar{l}_4}{16\pi^2}
+24L^{\mathrm{PQ,r}}_0+8L^{\mathrm{PQ,r}}_3-8L^{\mathrm{PQ,r}}
_5\right.\right.\nonumber\\
&&\left.\left.+8L^{\mathrm{PQ,r}}_8+\frac{1}{64\pi^2}\right)\right],\nonumber\\
a_C^{00}\ali =\ali \frac{M_\pi^2}{32\pi 
F_\pi^2}\left[1-2\mu_\pi+\frac{M_\pi^2}{F_\pi^2}\left(-48L^{\mathrm{PQ,r}}
_0-16L^{\mathrm{PQ,r}}_3+16L^{\mathrm{PQ,r}}_5-16L^{\mathrm{PQ,r}}_8-\frac{1}{
16\pi^2}\right)\right],\nonumber\\
a_R^{00}\ali =\ali \frac{3M_\pi^2}{16\pi 
F_\pi^2}\left[1-2\mu_\pi+\frac{M_\pi^2}{F_\pi^2}\left(48L^{\mathrm{PQ,r}}_0+16L^
{\mathrm{PQ,r}}_3+16L^{\mathrm{PQ,r}}_8+\frac{1}{8\pi^2}\right)\right],
\nonumber\\
a^{00}_V\ali =\ali \frac{9M_\pi^2}{16\pi 
F_\pi^2}\left[\mu_\pi+\frac{M_\pi^2}{F_\pi^2}\left(\frac{\bar{l}_1}{72\pi^2}
+\frac{\bar{l}_2}{36\pi^2}-\frac{\bar{l}_3}{192\pi^2}+\frac{\bar{l}_4}{16\pi^2}
-8L^{\mathrm{PQ,r}}_0-\frac{8L^{\mathrm{PQ,r}}_3}{3}-\frac{8L^{\mathrm{PQ,r}}_5}
{3}\right.\right.\nonumber\\
&&\left.\left.-\frac{8L^{\mathrm{PQ,r}}_8}{3}+\frac{5}{192\pi^2}\right)\right],
\nonumber\\
a^{00}_{\mathrm{tot}}\ali =\ali \frac{7M_\pi^2}{32\pi 
F_\pi^2}\left[1+\frac{M_\pi^2}{\pi^2F_\pi^2}\left(\frac{5\bar{l}_1}{84}+\frac{
5\bar{l}_2}{42}-\frac{5\bar{l}_3}{224}+\frac{\bar{l}_4}{8}+\frac{5}{32}
\right)\right];
\label{eq:a00}
\end{eqnarray}
\subsection*{$I=2,J=0$:}
\begin{eqnarray}
a^{20}_D\ali =\ali a^{00}_D \, ,\nonumber\\
a^{20}_C\ali =\ali -2a^{00}_C \, ,\nonumber\\
a^{20}_{\mathrm{tot}}\ali =\ali -\frac{M_\pi^2}{16\pi 
F_\pi^2}\left[1+\frac{M_\pi^2}{\pi^2F_\pi^2}\left(-\frac{\bar{l}_1}{12}-\frac{
\bar{l}_2}{6}+\frac{\bar{l}_3}{32}+\frac{\bar{l}_4}{8}-\frac{1}{32}\right)\right
];
\end{eqnarray}
\subsection*{$I=1,J=1$:}
\begin{eqnarray}
a^{11}_D\ali =\ali \frac{1}{12\pi 
F_\pi^2}\left[\mu_\pi+\frac{M_\pi^2}{F_\pi^2}\left(-\frac{\bar{l}_1}{24\pi^2}
+\frac{\bar{l}_2}{24\pi^2}+\frac{\bar{l}_4}{16\pi^2}+8L^{\mathrm{PQ,r}}_3-4L^{
\mathrm{PQ,r}}_5-\frac{13}{384\pi^2}\right)\right],\nonumber\\
a^{11}_R\ali =\ali \frac{1}{24\pi 
F_\pi^2}\left[1-2\mu_\pi+\frac{M_\pi^2}{F_\pi^2}\left(-16L^{\mathrm{PQ,r}}_3+8L^
{\mathrm{PQ,r}}_5-\frac{13}{288\pi^2}\right)\right],\nonumber\\
a^{11}_{\mathrm{tot}}\ali =\ali \frac{1}{24\pi 
F_\pi^2}\left[1+\frac{M_\pi^2}{\pi^2F_\pi^2}\left(-\frac{\bar{l}_1}{12}+\frac{
\bar{l}_2}{12}+\frac{\bar{l}_4}{8}-\frac{65}{576}\right)\right];
\end{eqnarray}
\subsection*{$I=0,J=2$:}
\begin{eqnarray}
a^{02}_D\ali =\ali \frac{1}{90\pi F_\pi^4}\left[\frac{F_\pi^2}{M_\pi^2}
\mu_\pi+\frac{\bar{l}_1}{16\pi^2}+\frac{\bar{l}_2}{16\pi^2}-
24\pi^2L^{\mathrm{PQ,r}}_0-12\pi^2L^{\mathrm{PQ,r}
}_3-\frac{77}{640\pi^2}\right],\nonumber\\
a^{02}_C\ali =\ali \frac{1}{180\pi F_\pi^4}\left[\frac{F_\pi^2}{M_\pi^2}
\mu_\pi- 24\pi^2L^{\mathrm{PQ,r}}_0-
12\pi^2L^{\mathrm{PQ,r}}_3+\frac{13}{320\pi^2} \right],\nonumber\\
a^{02}_R\ali =\ali \frac{1}{30\pi F_\pi^4}\left[-\frac{F_\pi^2}{M_\pi^2}
\mu_\pi+24L^{\mathrm{PQ,r}}_0+6L^{\mathrm{PQ,r}}_3-\frac{57}{1920\pi^2}
\right],\nonumber\\
a^{02}_V\ali =\ali \frac{1}{60\pi F_\pi^4}\left[\frac{F_\pi^2}{M_\pi^2}
\mu_\pi+ \frac{\bar{l}_2}{8\pi^2}-24 L^{\mathrm{PQ,r}}_0
-\frac{3}{20\pi^2}\right],\nonumber\\
a^{02}_{\mathrm{tot}}\ali =\ali \frac{1}{1440\pi^3F_\pi^4}\left[\bar{l}_1+4\bar{l}
_2-\frac{53}{8}\right];
\label{eq:a02}
\end{eqnarray}
\subsection*{$I=2,J=2$:}
\begin{eqnarray}
a^{22}_D\ali =\ali a^{02}_D \, ,\nonumber\\
a^{22}_C\ali =\ali -2a^{02}_C \, ,\nonumber\\
a^{22}_{\mathrm{tot}}\ali =\ali \frac{1}{1440\pi^3F_\pi^4}
\left[\bar{l}_1+\bar{l}_2-\frac{103}{40}\right].
\end{eqnarray}

\bigskip

\end{appendix}

\bibliographystyle{elsarticle-num} 
\bibliography{PQxPT_ref}

\end{document}